\documentclass[11pt]{article}
\usepackage[utf8]{inputenc}
\usepackage[T1]{fontenc}
\usepackage{lmodern}
\usepackage{geometry}
\usepackage{graphicx}
\usepackage[dvipsnames]{xcolor}
\usepackage[colorlinks=true]{hyperref}
\hypersetup{
    linkcolor=Magenta,
    citecolor=ForestGreen,
    urlcolor=Magenta
}
\usepackage{amsmath,amssymb,amsthm,amsfonts,latexsym,bbm,xspace,graphicx,float,mathtools,braket}
\usepackage{algorithm}
\usepackage{algpseudocode}
\usepackage{complexity}
\usepackage{appendix}
\usepackage{enumitem}
\usepackage[nameinlink,capitalize]{cleveref}  
\theoremstyle{plain}
\newtheorem{theorem}{Theorem}[section]
\newtheorem{lemma}[theorem]{Lemma}
\newtheorem{corollary}[theorem]{Corollary}

\renewcommand{\hat}[1]{\widehat{#1}}
\newcommand{\abs}[1]{|#1|}
\newcounter{question}
\newcommand{\question}[1]{%
    \par\medskip
    \refstepcounter{question}%
    \noindent\textbf{Question \thequestion.} #1
    \par\medskip
}

\renewcommand{\poly}{\operatorname{poly}}
\let\eps\varepsilon

\allowdisplaybreaks

\title{State Synthesis with an NP oracle}
\author{
  Aniruddha Sen\thanks{aniruddhasen@utexas.edu} \vspace{0.3em} \\ \emph{The University of Texas at Austin}
}

\date{November 1, 2026}

\begin{document}

\maketitle

\begin{abstract}
In this note, I describe a quantum algorithm that prepares an arbitrary state $\ket{\psi}$ to inverse exponential precision in polynomial time by making queries to an $\NP$ oracle, assuming one can directly query entries of the state vector corresponding to $\ket{\psi}$. The previous best upper bound on the complexity class required by an oracle for this task was $\PP$. I provide evidence that $\NP$ is the best possible upper bound by showing that any quantum algorithm using a $\P$ oracle, or even a $\Ppoly$ oracle, requires exponentially many queries. I also discuss some implications of this result for the $\QMA$ search-to-decision problem and unitary synthesis.
\end{abstract}

\section{Introduction}
The state synthesis problem was first studied by Aaronson \cite{aaronson2016complexity}, and asks how many queries a quantum algorithm $\mathcal A$ needs to make to some classical boolean function $f$ such that $\mathcal{A}^f$ approximately prepares an arbitrary state $\ket{\psi}$. The circuit corresponding to $\mathcal{A}$ should be the same for every state, but the function $f$ is allowed to arbitrarily depend on $\ket{\psi}$. This problem is motivated by asking if the complexity of preparing any state can be \emph{reduced} to the complexity of implementing some classical boolean function. A natural question to ask is then how weak can we make $f$, in terms of its complexity, such that we could efficiently prepare an arbitrary $\ket{\psi}$? This is the central question considered in this note. Let us assume a \emph{meta-oracle} $O_{\ket{\psi}}$ (or just $O$) can directly query entries of $\ket{\psi}$ to any desired bits of precision, since in general the amplitudes may even be uncomputable. The main result is as follows. 

\begin{theorem}\label{thm:one}
    There exists a quantum algorithm $\mathcal{A}$ running in $\poly(n,\log(1/\eps))$ time such that, for every $n$-qubit state $\ket{\psi}$, there exists an oracle $f \in \NP$, such that $\mathcal{A}^{f^{O}}$ prepares a state $\ket{\hat{\psi}}$ having fidelity at least $1-\eps$ with $\ket{\psi}$. The algorithm $\mathcal{A}$ makes $\poly(n)$ queries to $f^O$, where $O$ is an oracle for the entries of $\ket{\psi}$.
\end{theorem}

Setting $\eps=2^{-n}$ implies $\mathcal{A}$ can prepare $\ket{\psi}$ to inverse-exponential error in $\poly(n)$ time. Furthermore, a second result shows that if $f$ is allowed to be in $\P^\NP$ rather than just $\NP$, the algorithm requires at most $O(n/\log n)$ queries. This is proved in \cref{sec:alg}. Now, what was previously known about the power of $f$ needed? The best known upper bound was that $f \in \PP$ suffices, initially by an algorithm using $O(n)$ queries \cite{aaronson2016complexity}. Prior work on state synthesis has mainly focused on reducing the number of queries needed for this task. In particular, \cite{irani2021quantum} showed that a $\PP$ oracle with 1 query suffices for state synthesis to obtain inverse-polynomial precision, which was improved by Rosenthal, showing 1 query suffices to obtain inverse-exponential precision \cite{rosenthal2024efficient}. However, the construction in \cite{rosenthal2024efficient} uses an even stronger $\PSPACE$ oracle. In this work, I will sacrifice the progress made towards reducing the number of queries and just focus on reducing the power of the oracle instead. The algorithm presented here builds on the original state synthesis algorithm by Aaronson \cite{aaronson2016complexity}, from which one can notice it almost directly follows that $f \in \NP$ suffices to get inverse-polynomial error, and which forms the starting point of the algorithm here. The surprising part is that we can get inverse-exponential error as well. The key ideas combine simple insights from complexity theory and quantum algorithms, in particular including \emph{approximate counting with an NP oracle} \cite{stockmeyer1985approximation}, \emph{quantum rejection sampling} \cite{ozols2013quantum}, and \emph{exact amplitude amplification} \cite{brassard2000quantum}. The algorithm is described in \cref{sec:alg}. I provide evidence that it is optimal in the sense of $\NP^O$ being the weakest oracle possible for efficient state synthesis, by showing in the \hyperref[app]{Appendix} that a $\P^O$ oracle does not suffice, even when given polynomially bits of advice. The proof goes through a straightforward reduction from the lower bound for unordered search. To motivate this result further, let us consider the following direct consequence of \cref{thm:one}.

\begin{corollary}\label{cor:one}
    Consider a language $L$ such that $\NP^L \subseteq \BQP^L$. Then, any quantum state $\ket{\psi}$ whose amplitudes can be computed in polynomial time with an oracle for $L$ can also be prepared to inverse-exponential error in quantum polynomial time with an oracle for $L$.
\end{corollary}

This follows just by observing that for such a language, $f \in \NP^{P^L}=\NP^L \subseteq \BQP^L$ suffices to apply to prepare $\ket{\psi}$ by applying \cref{thm:one}. This shows that the search problem of preparing a state reduces to the problem of computing the amplitudes, given an $\NP^L$ oracle can be implemented efficiently. Notice that inverse-exponential error is in some sense the closest we can get to exact preparation, since beyond this we run into issues that involve specifying a gate set. Most interesting applications require this error guarantee in state preparation; it has been important in the context of state synthesis \cite{irani2021quantum,rosenthal2024efficient}, as well as in defining unitary complexity classes \cite{kretschmer2026unitary}. Now, it is well known that state synthesis is the quantum version of classically sampling from a probability distribution. Consequently, \cref{thm:one} can also be viewed as a quantum state witness analogue of classical results by Bellare, Goldreich, Petrank \cite{bellare2000uniform} and Jerrum, Valiant, Vazirani \cite{jerrum1986random}, who show how to exactly sample $\NP$-witnesses given access to an $\NP$ oracle. Indeed, the techniques here bear similarities to the approach by \cite{jerrum1986random}. Further discussion and open problems are deferred to \cref{sec:disc}.

\section{The Algorithm}\label{sec:alg}
I restate the main theorem here for convenience, followed by its proof.
\begin{theorem}[Restatement of \cref{thm:one}]\label{thm:one_sec2}
    There exists a quantum algorithm $\mathcal{A}$ running in $\poly(n, \log(1/\eps))$ time such that, for every $n$-qubit state $\ket{\psi}$, there exists an oracle $f \in \NP$, such that $\mathcal{A}^{f^{O}}$ prepares a state $\ket{\hat{\psi}}$ having fidelity at least $1-\eps$ with $\ket{\psi}$. The algorithm $\mathcal{A}$ makes $\poly(n)$ queries to $f^O$, where $O$ is an oracle for the entries of $\ket{\psi}$.
\end{theorem}
\begin{proof}
Our starting point is an algorithm first proposed by Grover and Rudolph \cite{grover208112creating} for preparing quantum states and adapted in the context of state synthesis by Aaronson \cite{aaronson2016complexity}. We first briefly describe this algorithm for clarity, setting the desired error $\eps=O(2^{-n})$ for ease of exposition. It involves preparing the target state $\ket{\psi}$ by computing \emph{prefix-sums}. Consider the probability distribution $\{p_x\}$ over $n$-length bitstrings $x \in \{0,1\}^n$ corresponding to the amplitudes of $\ket{\psi}$ up to polynomial (or $\mathrm{polylog}(1/\eps)=\poly(n)$) bits of precision. Let us define the prefix sum $s_{y} \coloneq \sum_{z} p_{yz}$, where $z$ consists of all $(n-\abs{y})$ length bitstrings. In other words, $s_{y}$ is the marginal probability when fixing the prefix to be $y$. We start by preparing the 1-prefix state $ \ket{\psi_1} \coloneq  (\sqrt{s_0}\ket{0} + \sqrt{s_1}\ket{1})\otimes \ket{0}$, followed by applying a rotation on the second qubit conditioned on the value of the first qubit such that it prepares the 2-prefix state as follows. 
\begin{align}
    \ket{\psi_2} &= \sqrt{s_0}\ket{0}\left(\sqrt{\frac{s_{00}}{s_0}}\ket{0} + \sqrt{\frac{s_{01}}{s_0}}\ket{1}\right) + \sqrt{s_1}\ket{1}\left(\sqrt{\frac{s_{10}}{s_1}}\ket{0} + \sqrt{\frac{s_{11}}{s_1}}\ket{1}\right) \\
    &= \sqrt{s_{00}}\ket{00} + \sqrt{s_{01}}\ket{01} + \sqrt{s_{10}}\ket{10} + \sqrt{s_{11}}\ket{11} \,.
\end{align}
This involves querying the oracle $f$ in order to compute the value of each conditional rotation depending on $s_y$, followed by a unitary that implements it, and then uncomputing the rotation value at the end with another query. We iteratively apply this process of conditional rotations, preparing each $k$-prefix state, which is defined as follows.
\begin{equation}
    \ket{\psi_k} \coloneq \sum_{y \in \{0,1\}^k} \sqrt{s_{y}}\ket{y}  \,.
\end{equation}
After repeating this process $n$ times, we obtain the so-called \emph{QSample} state $\ket{\psi_n}$. To reduce the number of queries for this, we can instead also \emph{batch} the iteration into groups of $\log (n)$ qubits. Each controlled rotation just gets replaced by some unitary on $\log(n)$ qubits, which remains efficient to implement. But the number of queries reduces to $O(n/\log n)$. Now, in order to prepare $\ket{\psi}$, we only need to add the phases $\{e^{i\theta_x}\}$, which we can do with just 2 extra queries to $f$: query the phase on an ancilla register, apply a conditional phase oracle, and finally uncompute the phase. Since each prefix sum requires adding an exponential list of non-negative real numbers, this algorithm requires $f$ to be a $\PP$ oracle.

Our goal will now be to relax the requirement for the $\PP$ oracle to an $\NP$ oracle. The first key observation we need to make for our algorithm is that approximate counting \cite{stockmeyer1985approximation} allows us to estimate counting problems in $\#\P$ to inverse polynomial precision in $\poly(n)$ time with a $\BPP^\NP$ oracle. Since our quantum algorithm can simulate the randomness required by any $\BPP$ algorithm, this shows that we can calculate any prefix sum to inverse polynomial precision. A simple calculation shows that this allows us to prepare $\ket{\psi}$ to inverse polynomial precision. However, our goal is to prepare $\ket{\psi}$ to inverse-exponential precision, as discussed in the introduction. Then, it seems like we made almost no progress, since an $\NP$ oracle does not let us calculate prefix sums to inverse-exponential precision, unless $\P^\NP=\P^\PP$.

This brings us to our second key idea. We will use the state prepared by approximate counting as a \emph{warm start} for the rest of our algorithm. We will invoke our $\NP$ oracle again, importantly making use of the fact that an $\NP$ oracle can trivially access each probability amplitude of $\ket{\psi}$ in superposition. This suggests we could try to check how close each approximation was and try to factor out the error, classically known as \emph{rejection sampling}. This leads us to utilizing a slightly modified version of \emph{quantum} rejection sampling---first proposed by Ozols, Roetteler, and Roland \cite{ozols2013quantum}---as follows.

We start with the state $\ket{\psi_{\mathrm{apx}}}$ as prepared through approximate counting of prefix-sums followed by the appropriate phase queries. Then, we again use our $\NP$ oracle to query the true probability amplitude $p_x$ of each basis state\footnote{\label{fn}up to only $\mathrm{polylog}(1/\eps)=\poly(n)$ bits of precision, so note that $\sum p_x$ may not equal 1, as $p_x$ is not normalized. In fact, calculating this normalization itself is in $\#\P$. However, we assume it is normalized for now, and account for it only in the error analysis. This will just incur an additional inverse-exponential error.} onto an additional ancilla register in superposition. We also query the \emph{value} of $q_x$ onto yet another register, which just requires us to repeat our initial $\NP$ queries over the same choice of randomness. So, we obtain the state
\begin{equation}
\ket{\psi_{\mathrm{apx}}'}=\sum_{x}\sqrt{q_x}e^{i\theta_x}\ket{x}\ket{p_x}\ket{q_x} \,,
\end{equation}
where each $q_x$ is obtained through a product of sufficiently good approximate prefix-sums such that it satisfies the guarantee that $\sqrt{p_x/2} \le \sqrt{q_x} \le \sqrt{2p_x}$. Next, we apply a rotation gate on one more ancilla register, controlled on the value of $p_x,q_x$ on the second and third register, such that (after uncomputing and discarding the two intermediate registers) it prepares the following state $\ket{\psi_{\mathrm{apx}}''}$. The guarantee on the ratio between $p_x$ and $q_x$ is crucial in order for this rotation to be \emph{both} approximately balanced and well-defined, which shows why our warm start state through approximate counting is needed for our algorithm.
\begin{align}
\ket{\psi_{\mathrm{apx}}''}&=\sum_{x}\sqrt{q_x}e^{i\theta_x}\ket{x} \otimes \left(\frac{\sqrt{2q_x-p_x}}{\sqrt{2q_x}}\ket{0} + \frac{\sqrt{p_x}}{\sqrt{2q_x}}\ket{1}\right) \, \\
&= \left(\frac{1}{\sqrt{2}}\sum_x \sqrt{2q_x-p_x}e^{i\theta_x}\ket{x}\right)\otimes\ket{0} + \left(\frac{1}{\sqrt{2}}\sum_x \sqrt{p_x}e^{i\theta_x}\ket{x}\right)\otimes\ket{1} \,.
\end{align}
Note that $\ket{\psi_{\mathrm{apx}}''}$ post-selected on measuring $\ket{1}$ on the second register collapses to the state $\ket{\psi}$. So it only remains to boost the probability of measuring $\ket{1}$, which we will call the \emph{good} branch of our superposition. Thus, the final subroutine we need is amplitude amplification.  Notice that the probability on the good branch is $\approx 1/2$, since $\sum_{x} p_x \approx 1$ up to an inverse-exponential error (see footnote~\ref{fn}). So, we just run an exact amplitude amplification algorithm \cite{brassard2000quantum} to boost the probability of the $\ket{1}$ branch from $\approx 1/2$ to $\approx 1$. Then, we simply measure and discard the second register. Thus, as desired, we have prepared the state 
\begin{align}
    \ket{\hat{\psi}} \coloneq \sum_x \sqrt{p_x}e^{i\theta_x}\ket{x} \,,
\end{align}
such that $\abs{\langle\hat\psi\ket{\psi}}^2 \ge 1-\eps$, given some $\eps > 0$. The error guarantee and runtime of this algorithm are proved in \cref{lem:two} as follows.
\end{proof}
\begin{lemma}\label[lemma]{lem:two}
The algorithm in \cref{thm:one} makes at most $\poly(n)$ queries to an $\NP$ oracle---alternately $O(n/\log n)$ queries to a $\P^\NP$ oracle---runs in at most $\poly(n,\log(1/\eps))$ time, and outputs a state $\ket{\hat{\psi}}$ such that $\abs{\langle\hat\psi\ket{\psi}}^2 \ge 1-\eps$.   
\end{lemma} 
\begin{proof}
Preparing the approximate counting state and querying the values of approximate counts $q_x$ dominates the number of queries required. In particular, we need $O(n/\log n)$ queries to an approximate counting subroutine. In addition, we just require a constant number of additional $\NP$ queries to get the phases and the values of $p_x$ (to $\textrm{polylog}(1/\eps))$ bits) onto an ancilla register, via the Bernstein--Vazirani trick \cite{irani2021quantum}. Exact amplitude amplification requires us to repeat this process a constant number of times. Thus, given a $\P^\NP$ oracle, our algorithm would need just $O(n/\log n)$ queries, as it could give us the approximate counting values directly \cite{stockmeyer1985approximation} given access to random bits. To verify we get the desired error guarantee, recall that we require each probability amplitude $q_x$ to be within a factor of 2 of $p_x$. Each $q_x$ is a product of $O(n/\log n)$ ratios of approximate prefix sums, so estimating every prefix sum to within a factor $(1\pm\gamma)$ with $\gamma = \Theta(\log n/n)$ guarantees this; a $\BPP^{\NP}$ algorithm just requires $\poly(n,1/\gamma,\log(1/\eps))=\poly(n,\log(1/\eps))$ time for this, with failure probability $\eps$ after a union bound and boosting through repetition \cite{stockmeyer1985approximation}.

We can simulate each such query to a $\BPP^\NP$ computation with $\poly(n)$ queries of length $\poly(n,\log(1/\eps))$ to an $\NP$ oracle, over a fixed choice of random bits throughout. Thus, the algorithm requires polynomial queries; and since each unitary we implement takes at most $\poly(n,\log(1/\eps))$ time (to implement any controlled rotation for instance), the algorithm also runs in $\poly(n,\log(1/\eps))$ time.

Now, let us bound the error in preparing the state given the above error guarantee on each prefix sum. With this guarantee, we know that the rotation gate to prepare the state $\ket{\psi_{\mathrm{apx}}''}$ is well-defined. Now, we just bound the error between $\ket{\psi_{\mathrm{apx}}''}$ and the \emph{ideal} such state $\ket{\psi''}$ that has the exact $p_x,\theta_x,q_x$ encoded in its amplitudes. Each entry of their state vectors agrees up to $O(\sqrt{\eps})$ relative error, so the two states are within $O(\sqrt{\eps})$ in $L^2$ norm. After constant rounds of amplitude amplification, note that $\ket{\psi''}$ becomes exactly $\ket{\psi}$ and for $\ket{\psi_{\mathrm{apx}}''}$, amplitude amplification uses unitaries that are at most within $O(\sqrt{\eps})$ operator norm of the ideal ones, so we end up getting a state within $O(\sqrt{\eps})$ of $\ket{\psi}$. Thus, except with probability $\eps$ over the random bits, $\abs{\langle\hat\psi\ket{\psi}}^2 \ge 1-O(\eps)$. So, by a standard argument, we can prepare $\ket{\hat{\psi}}$ (with the same global phase) in superposition over all choices of $\poly(n,\log(1/\eps))$ size random bits, and then cleanly uncompute them at the end, along with rescaling $\eps$, to deterministically obtain $\ket{\hat{\psi}}$ such that $\abs{\langle\hat\psi\ket{\psi}}^2 \ge 1-\eps$.
\end{proof}

\section{Discussion}\label{sec:disc}
This algorithm highlights the power of $\BQP$ with an $\NP$ oracle, addressing a question raised explicitly in \cite{gharibian2024bqp}. This result also extends a result by Grover and Rudolph \cite{grover208112creating}, who showed how to prepare quantum states corresponding to efficiently \emph{integrable} distributions; here we just require the distribution to be \emph{approximately} integrable---to $1/\poly(n)$ error instead of $1/\exp(n)$---as long as it is also pointwise efficiently \emph{computable}. Further, slightly modifying the algorithm also approximately prepares, without the oracle $O$, any state with amplitudes computable in $\#\P$. 

Let us now consider the longstanding question of whether $\QMA$-complete problems admit search-to-decision reductions. \cref{thm:one} implies that $\QMA$-search, which is the problem of preparing $\QMA$ witness states, essentially reduces to the complexity of computing the amplitudes of the state, since $\NP \subseteq \QMA$.  However, it seems unlikely the amplitudes themselves could also be computed in $\QMA$, since the amplitudes could encode a $\GapP$-complete problem. Considering the contrapositive direction might be more useful. If we want to prove that there exists a classical oracle $C$ relative to which search-to-decision does not exist for $\QMA$, the oracle needs to satisfy the property that amplitudes of $\QMA^C$ witness states cannot be computed in any class $L$ such that $\NP^{(L,C)} \subseteq \QMA^{C}$.

\paragraph{Open Problems.}
There are a few concrete future directions that appear interesting. The first is motivated by the original question posed by \cite{aaronson2016complexity} on reducing the number of queries for state synthesis.   
\question{Can the number of queries for efficient state synthesis with an $\NP$ oracle be made sublinear or even constant?}
I conjecture that the answer to this question is yes, perhaps requiring only one query, and that it would need to go through a technique that does not involve computing prefix-sums, since this seems inherently limited at $O(n/\log n)$ queries, even with a $\P^\NP$ oracle. I believe the approach given by the algorithm here could extend to such an upper bound with just an $\NP$ oracle, via a more fine-grained analysis of the approximate counting step in particular, but no further. Weakening the oracle required by constant query state synthesis algorithms \cite{irani2021quantum,rosenthal2024efficient} appears to be the obvious initial route towards this goal.

There is also a broader question we could consider. The Unitary Synthesis Problem asks if any unitary can be approximately implemented by a quantum algorithm making a polynomial number of queries to a classical boolean function \cite{aaronson2016complexity,lombardi2024one}. It is clear that state synthesis is a special case of unitary synthesis. Thus, \cref{thm:one} directly prompts the following question.
\question{For every $n$-qubit unitary $U$, does there exist a function $f \in \NP$, such that $U$ can be approximately implemented by a quantum algorithm making a polynomial number of queries to an $f^O$ oracle?}
Here, the meta-oracle $O$ would have query access to the entries of the classical description of $U$ in the computational basis. I conjecture that the answer to this question is also yes, largely due to a recent claimed solution by OpenAI for the Unitary Synthesis Problem \cite{OAI}, and related progress by Nehoran, Slote, and Yuen \cite{nehoran2026all}. A clear route to proving this would be to reduce an algorithm for unitary synthesis to just using state synthesis as a black box, with some slack of additional queries that require nothing more powerful than an $\NP^O$ oracle.  Conversely, it is possible that $\NP$ may be the weakest non-trivial class (i.e., larger than $\P$) with respect to which an exponential query lower bound exists. Proving this would likely require new techniques, since the lower bound would need to be with respect to a specific complexity of the oracle, rather than an arbitrary boolean function.

\section*{Acknowledgments}
I thank Scott Aaronson, William Kretschmer, and Ronak Ramachandran for helpful discussions and comments. Generative AI was used for literature search, assisting with the error analysis, and proofreading the final draft.

\bibliographystyle{alpha}
\bibliography{references}

@article{grover208112creating,
  title={Creating superpositions that correspond to efficiently integrable probability distributions},
  year={2002},
  author={Grover, L and Rudolph, T},
  journal={arXiv preprint quant-ph/0208112}
}

@article{ozols2013quantum,
  title={Quantum rejection sampling},
  author={Ozols, Maris and Roetteler, Martin and Roland, J{\'e}r{\'e}mie},
  journal={ACM Transactions on Computation Theory (TOCT)},
  volume={5},
  number={3},
  pages={1--33},
  year={2013},
  publisher={ACM New York, NY, USA}
}

@article{stockmeyer1985approximation,
  title={On approximation algorithms for\# P},
  author={Stockmeyer, Larry},
  journal={SIAM Journal on Computing},
  volume={14},
  number={4},
  pages={849--861},
  year={1985},
  publisher={SIAM}
}

@article{aaronson2016complexity,
  title={The complexity of quantum states and transformations: from quantum money to black holes},
  author={Aaronson, Scott},
  journal={arXiv preprint arXiv:1607.05256},
  year={2016}
}

@inproceedings{rosenthal2024efficient,
  title={Efficient quantum state synthesis with one query},
  author={Rosenthal, Gregory},
  booktitle={Proceedings of the 2024 Annual ACM-SIAM Symposium on Discrete Algorithms (SODA)},
  pages={2508--2534},
  year={2024},
  organization={SIAM}
}

@article{irani2021quantum,
  title={Quantum search-to-decision reductions and the state synthesis problem},
  author={Irani, Sandy and Natarajan, Anand and Nirkhe, Chinmay and Rao, Sujit and Yuen, Henry},
  journal={arXiv preprint arXiv:2111.02999},
  year={2021}
}

@article{brassard2000quantum,
  title={Quantum amplitude amplification and estimation},
  author={Brassard, Gilles and Hoyer, Peter and Mosca, Michele and Tapp, Alain},
  journal={arXiv preprint quant-ph/0005055},
  year={2000}
}

@article{kretschmer2026unitary,
  title={Unitary complexity in polynomial space},
  author={Kretschmer, William and Tang, Ewin},
  journal={arXiv preprint arXiv:2610.03705},
  year={2026}
}

@article{bellare2000uniform,
  title={Uniform generation of NP-witnesses using an NP-oracle},
  author={Bellare, Mihir and Goldreich, Oded and Petrank, Erez},
  journal={Information and Computation},
  volume={163},
  number={2},
  pages={510--526},
  year={2000},
  publisher={Elsevier}
}

@article{jerrum1986random,
  title={Random generation of combinatorial structures from a uniform distribution},
  author={Jerrum, Mark R and Valiant, Leslie G and Vazirani, Vijay V},
  journal={Theoretical computer science},
  volume={43},
  pages={169--188},
  year={1986},
  publisher={Elsevier}
}

@article{nayebi2014quantum,
  title={Quantum lower bound for inverting a permutation with advice},
  author={Nayebi, Aran and Aaronson, Scott and Belovs, Aleksandrs and Trevisan, Luca},
  journal={arXiv preprint arXiv:1408.3193},
  year={2014}
}

@article{bennett1997strengths,
  title={Strengths and weaknesses of quantum computing},
  author={Bennett, Charles H and Bernstein, Ethan and Brassard, Gilles and Vazirani, Umesh},
  journal={SIAM journal on Computing},
  volume={26},
  number={5},
  pages={1510--1523},
  year={1997},
  publisher={SIAM}
}

@article{gharibian2024bqp,
  title={BQP, meet NP: Search-to-decision reductions and approximate counting},
  author={Gharibian, Sevag and Kamminga, Jonas},
  journal={arXiv preprint arXiv:2401.03943},
  year={2024}
}

@inproceedings{lombardi2024one,
  title={A one-query lower bound for unitary synthesis and breaking quantum cryptography},
  author={Lombardi, Alex and Ma, Fermi and Wright, John},
  booktitle={Proceedings of the 56th Annual ACM Symposium on Theory of Computing},
  pages={979--990},
  year={2024}
}

@misc{OAI,
  key = {{OAI}},
  title = {{OpenAI. Polynomial-Time Unitary Synthesis from a Boolean Oracle}},
  howpublished = {OpenAI Math Release preprint github.com/\allowbreak openai/\allowbreak math/\allowbreak blob/\allowbreak main/\allowbreak preprints/\allowbreak Polynomial-Time-Unitary-Synthesis-from-a-Boolean-Oracle-October-5-2026/\allowbreak paper.pdf},
  year = {2026}
}

@article{nehoran2026all,
  title={All Unitaries Have Constant Depth Quantum Circuits},
  author={Nehoran, Barak and Slote, Joseph and Yuen, Henry},
  journal={arXiv preprint arXiv:2609.40351},
  year={2026}
}

\appendix

\section{Lower bound with a $\Ppoly^O$ oracle}\label{app}
\begin{lemma}
Consider a quantum algorithm $\mathcal{A}$ such that, for every $n$-qubit state $\ket{\psi}$, there exists an oracle $f \in \Ppoly$, such that $\mathcal{A}^{f^O}$ prepares a state $\ket{\hat{\psi}}$ having at least $\Omega(1)$ fidelity with $\ket{\psi}$. Then, $\mathcal{A}$ needs to make $\Tilde{\Omega}(\sqrt{2^n})$ queries to $f^O$, where $O$ is an oracle for the entries of $\ket{\psi}$.
\end{lemma}
\begin{proof}
We give a simple reduction from the lower bound for unordered search. Consider the set of basis states $\{\ket{x}\}$, $x \in \{0,1\}^n$, and consider the simpler problem where the algorithm $\mathcal{A}$ just needs to prepare some basis state $\ket{i}$ and is allowed to make queries to $f$, but $f$ only has query access to the entries of $i$. Since $f \in \Ppoly$, $\BQP$ can simulate any queries $f$ makes to $O$ directly with polynomial bits of advice. Now, notice that any algorithm $\mathcal{A}^O$ that could prepare $\ket{i}$ would solve the search problem for finding a marked element with query access via $O$ to the $2^n$ size state vector of $\ket{\psi}$. By the lower bound for unordered search \cite{bennett1997strengths}, even with polynomial bits of advice \cite{nayebi2014quantum} (note that the advice can only depend on $n$), $\mathcal{A}$ needs to make at least $\Tilde{\Omega}(\sqrt{2^n})$ queries to $O$. This concludes the proof.
\end{proof}

\end{document}